\documentclass[12pt]{article}

\usepackage{amsmath}
\usepackage{amsfonts}
\usepackage{amssymb}
\def\hybrid{\topmargin -20pt    \oddsidemargin 0pt
        \headheight 0pt \headsep 0pt
        \textwidth 6.25in       
        \textheight 9.5in       
        \marginparwidth .875in
        \parskip 5pt plus 1pt   \jot = 1.5ex}
\hybrid
\numberwithin{equation}{section}
\numberwithin{table}{section}
\allowdisplaybreaks
\newcommand{\beq}{\begin{equation}}
\newcommand{\eeq}{\end{equation}}
\newcommand{\bea}{\begin{eqnarray}}
\newcommand{\eea}{\end{eqnarray}}
\newcommand{\ba}{\begin{array}}
\newcommand{\ea}{\end{array}}
\newcommand{\bt}{\begin{tabular}}
\newcommand{\et}{\end{tabular}}
\newcommand{\bc}{\begin{center}}
\newcommand{\ec}{\end{center}}

\newcommand{\ax}{\alpha}
\newcommand{\bx}{\beta}

\newcommand{\dx}{\delta}
\newcommand{\ox}{\omega}

\newcommand{\Ox}{\Omega}

\newcommand{\cL}{\mathcal{L}}
\newcommand{\cK}{\mathcal{K}}
\newcommand{\cN}{\mathcal{N}}

\newcommand{\cF}{\mathcal{F}}

\newcommand{\cM}{\mathcal M}

\newcommand{\wg}{\wedge}

\DeclareMathOperator{\SU}{\mathit{SU}}

\newcommand{\CY}{Calabi--Yau\hspace{0.2cm}}

\newcommand{\nn}{\nonumber}

\def\theequation{\arabic{section}.\arabic{equation}}

\newcommand{\tox}{\tilde\omega}
\newcommand{\txi}{\tilde\xi}
\newcommand{\IM}{\textrm{Im} \,}
\newcommand{\RE}{\textrm{Re} \,}

\newcommand{\moo}{\kappa_0}

\def\rme{{\rm e}}

\begin{document}

\begin{titlepage}
\begin{center}

\hfill hep-th/0212278\\

\vfill
{\large \bf  Type IIB Theory on Half-flat Manifolds}\footnote{%
Work supported by: the CNRS -- the French Center for National Scientific
Research,
DFG -- The German Science Foundation
and  DAAD -- the German Academic Exchange Service. }\\

\vskip 0.8cm

{\bf Sebastien Gurrieri}\\

\vskip 0.3cm
{\em Centre de Physique Th\'eorique, CNRS Luminy, Case 907,\\
F-13288 Marseille -Cedex 9, France}\\
{\tt gurrieri@cpt.univ-mrs.fr}
\vskip 0.8cm
{\bf Andrei Micu\footnote{On leave from IFIN-HH Bucharest}
}  \\

\vskip 0.3cm
{\em Fachbereich Physik, Martin-Luther-Universit\"at Halle-Wittenberg,\\
Friedemann-Bach-Platz 6, D-06099 Halle, Germany}\\
{\tt micu@physik.uni-halle.de}

\end{center}

\vskip 1cm

\begin{center} {\bf ABSTRACT } \end{center}

\noindent
In this note we derive the low-energy effective action of type IIB theory
compactified on half-flat manifolds and we show that this
precisely coincides with the low-energy effective action of type IIA theory
compactified on a \CY manifold in the presence of NS three-form fluxes. We
provide in this way a further check of the recently formulated conjecture that
half-flat manifolds appear as mirror partners of
\CY manifolds when NS fluxes are turned on.
\vfill

\noindent December 2002

\end{titlepage}


\section{Introduction}
\setcounter{equation}{0}

\CY compactification is one of the most common
procedures to obtain four-dimensional models from ten-dimensional string
theories. However, the physics obtained in this way generically feature a
large
number of scalar fields (moduli) which are flat directions of the potential
and moreover, there exist no viable mechanism to further break supersymmetry.
It was recently realized
\cite{Bachas}\,--\,\cite{GLMW} that if one allows a non-zero
background value (flux) for some of the field strengths, a
potential is generated in the lower-dimensional effective theory
and supersymmetry can be spontaneously broken.\footnote{Strictly
speaking this idea appeared for the first time in
  \cite{SS}. In the context of finding supersymmetric ground states
  this was initially addressed in \cite{AS1,Hull1,dWS}. More recently,
  orientifolds and Calabi--Yau four-folds with fluxes have been discussed in
\cite{BBHL,BD}, \cite{FP}\,--\,\cite{KSTT}.} Beside the
phenomenologically interesting features of such compactifications
it is also attractive to study flux compactifications in the
context of string dualities
\cite{JM}\,--\,\cite{CKKL}, \cite{CKL,LM2,GLMW,AAHV}, \cite{KSTT}\,--\,\cite{harvendra2}
and in particular, in this note we will concentrate on mirror symmetry which
is supposed to relate the two type II theories when compactified on mirror \CY
three-folds.

The issue of mirror symmetry when fluxes are turned on was
addressed in several works \cite{JM,TV,Vafa,CKL,LM2,GLMW,AAHV,KSTT}.
In type IIA theory the RR fluxes lie in the even cohomologies of
the \CY manifold as the RR sector of this theory contains even
form field strengths. For type IIB on the other hand one
encounters odd form field strengths and thus the RR fluxes are
parameterized in this case by elements of the odd cohomologies of
the \CY space. As mirror symmetry precisely exchanges the odd and
even cohomologies it is not surprising that it still holds when
RR fluxes are turned on. For the NS fluxes the situation was
until recently less clear as none of the two type II theories
contain even form field strengths in the NS-NS sector. It was in
turn proposed \cite{Vafa} that the mirror of the NS fluxes should
now come from the geometry of the internal manifold. This
proposal was made more concrete in \cite{GLMW} where it was
conjectured that when NS fluxes are turned on in type IIB theory,
mirror symmetry requires the presence of a new class of
manifolds, known as half-flat manifolds with
$\SU(3)$ structure\footnote{%
Manifolds with $\SU(3)$ structure also appeared recently in the
heterotic string compactifications \cite{CCDLMZ}, though from a
slightly different perspective.}
on type IIA side. The main argument supporting this
proposal was provided by showing that the low-energy effective actions for the
type IIB compactified on a \CY three-fold in the presence of electric NS
three-form flux and type IIA compactified on a half-flat space are
equivalent. The purpose of this note is to test the conjecture formulated in
\cite{GLMW} in the reversed situation, namely we want to show that
compactifying type IIB theory on half-flat manifolds produces an effective
action which is mirror equivalent to type IIA theory compactified on \CY
three-folds with NS three-form flux turned on.

The paper is structured as follows. In section~2 we briefly recall some of the
results obtained in \cite{GLMW} and mainly we are interested in
those features which are relevant for the KK reduction.
In section~3 we compute the low-energy effective action of type IIB
supergravity compactified
on such a manifold and show that it indeed reproduces the action obtained in
the type IIA case when NS fluxes are turned on. In section 4 we present our
conclusions while in the appendix we record the main steps of the
compactification of type IIA theory on \CY three-folds with NS fluxes
\cite{LM2}.
Throughout the paper we use the conventions of \cite{LM2} (see appendix A of
this paper).

\section{Preliminaries}


Let us start by recording the main results obtained in \cite{GLMW}. Turning on
NS three-form fluxes\footnotemark{} in type IIB compactification on a \CY
manifold $\tilde Y$ introduces $2(h^{(1,2)} +1)$ flux parameters via
\cite{JM}
\begin{equation}
  \label{NSflB}
  H_3 = p^A \ax_A + q_A \bx^A \ ,
\end{equation}
where $(\ax_A, \bx^B), \ A,B = 0, \ldots , h^{(1,2)}$ form a
basis for $H^3(\tilde Y)$ and is normalized as in \eqref{icy30}.
These fluxes will appear in the four
dimensional theory as charges which couple to electric or
magnetic fields and it is just pure convention to call them
electric or magnetic fluxes depending on how we choose to
describe the gauge sector. However, in the setup of \cite{LM2}
which we also adopt here, $p^A$ appear as magnetic charges, while
$q_A$ as electric ones. Thus, from now on we will refer to the
fluxes $p^A$ and $q_A$ in \eqref{NSflB} as magnetic and electric
fluxes respectively.

\footnotetext{By flux we understand a background value for some p-form field
  strength. This value can not be arbitrary as the equations of motion
  restrict it to be a harmonic form on the internal manifold. Thus we can
  write
  \begin{equation}
    F_{(p)} = m_i \; \ox^i_{(p)} \ , \nn
  \end{equation}
  where $m^i$ are the flux parameters and $\ox^i_{(p)}$ represents a basis of
  harmonic p-forms.}

The NS-NS sector of type IIA theory also contains a two form
potential with a three-form field strength. However, the
corresponding fluxes will again lie in the third cohomology and
they can not be mirror to \eqref{NSflB} since mirror symmetry
exchanges the even and odd cohomologies. In order to find a
configuration mirror symmetric to \eqref{NSflB} one needs to find
NS even-form field strengths.
It was suggested in \cite{Vafa} that
the missing fluxes should come from the geometry of the internal manifold
which now should be taken to be non-complex and
the NS even form field strength should be associated to the lack
of integrability of the almost complex structure.
It was shown in \cite{GLMW} that one can obtain the mirror electric fluxes by
considering the internal space to be a half-flat manifold with
$\SU(3)$ structure\footnote{%
For a systematic study of such manifolds we refer the reader to
\cite{salamon,CS} and references therein, or for a more physical discussion to
\cite{GLMW,CCDLMZ}.}
which is indeed non-complex.
Such manifolds admit a globally defined nowhere vanishing spinor which is
covariantly constant with respect to a connection with torsion
$\nabla^{(T)} \eta = 0$. This assures that the low-energy effective action
obtained by compactifying either of the type II theories on such manifolds
still has $N=2$ supersymmetries in four dimensions.
Equivalently, one can think about these spaces as being endowed with an almost
complex structure $J$ and a $(3,0)$ form $\Ox$ which are covariantly constant
with respect to the same connection with torsion
\begin{equation}
  \nabla_m^{(T)} J_{np} = 0 \ ; \qquad \nabla_m^{(T)} \Ox_{npq} = 0 \ .
\label{torsion1}
\end{equation}
The Levi-Civita connection fails to preserve $J$ and $\Ox$ thus, unlike the
\CY case $J$ and $\Ox$ are no longer closed. In \cite{GLMW} it was found that
the NS four-form was provided by $ d \Ox^+$, $\Ox^+$ being the real part of
the $(3,0)$ form $\Ox$, and the (electric) fluxes were obtained in the
expansion of this four-form in some appropriately chosen basis of $(2,2)$
forms $\tox^i$
\begin{equation}
  \label{elflA}
   d  \Ox^+ = e_i \tox^i \ .
\end{equation}
Moreover, it was argued that in order for mirror symmetry to work there should
be a precise relation between the half-flat space $\hat Y$ and some \CY
three-fold $Y$. In particular, the moduli space of metrics of the half-flat
manifold should coincide with the moduli space of the corresponding \CY and
the metrics on these spaces should be equivalent. This means that one should
perform the usual moduli expansion
\begin{equation}
  \label{modexp}
  \Ox = z^A \ax_A - \cF_A \bx^A \ , \qquad J = v^i \ox_i  \ ,
\end{equation}
and the forms $(\ax_A, \bx^B) $ and $(\ox_i, \tox^j)$ should have the same
intersection numbers as on the \CY manifold, i.e.
\begin{equation}
  \label{int}
  \int_{Y} \ax_A \wg \bx^B =  \dx_A^B \ , \qquad \int_Y \ox_i \wg \tox^j =
  \dx_i^j \ , \qquad \int_Y \ax_A \wg \ax_B = \int_Y \bx^A \wg \bx^B = 0 \ .
\end{equation}
The above relation proved to impose strong constraints on the
topology of the half-flat manifold $\hat Y$. In particular
it was argued in \cite{GLMW} that in order to have both \eqref{elflA} and
\eqref{modexp} the only solution is to consider the following
action of the exterior derivative on the basis $(\ax_A, \bx^B)$ \footnote{%
We have implicitly assumed that $z^A$ can be written as
$z^A = (1, z^a), \ a = 1, \ldots , h^{(1,2)}$.}
\begin{equation}
  \label{dax}
   d  \ax_0 = e_i \tox^i \ , \qquad  d  \ax_a =  d  \bx^B = 0 \ .
\end{equation}
Consistency with \eqref{int} further required that
\begin{equation}
  \label{dox}
   d  \ox_i = e_i \bx^0 \ , \qquad  d  \tox^i = 0 \ .
\end{equation}
Using these relations one can immediately see that the cohomology groups are
reduced compared to the corresponding \CY manifold and one has
\begin{equation}
  h^{(2)}(\hat Y) = h^{(1,1)}(Y) -1 \ , \qquad
  h^{(3)}(\hat Y) = h^{(3)}(Y) -2 \ .
\end{equation}
From a physical point of view this can be easily understood as,
due to the fluxes, some of the previous moduli of the \CY now
acquire masses and thus appear no longer as flat directions of the
potential. Consequently, in order to obtain the same light
spectrum some of the forms considered previously in the zero
modes expansion have to become non-harmonic on the mirror side.

Using the above setup it was shown in \cite{GLMW} that the
effective action of type IIA supergravity compactified on a
half-flat manifold $\hat Y$ precisely reproduces the effective
action of type IIB supergravity compactified on a \CY manifold
$Y$ in the presence of NS electric fluxes \eqref{NSflB} and thus
provides strong evidence that half-flat manifolds are indeed the
mirror configuration of the NS fluxes \eqref{NSflB}.

In this note we want to further test this conjecture. In particular, if the
half-flat geometries are to reproduce the mirror NS fluxes this should not
depend on which of the type II theories is chosen to be compactified on these
spaces. Thus, our
purpose here is to show that type IIB compactification on half-flat manifolds
reproduces the type IIA compactification on \CY three-folds in the presence
of electric NS fluxes whose effective action was derived in \cite{LM2}.

\section{Type IIB on half-flat manifolds}

Following \cite{GLMW} we will now perform the
compactification of type IIB on a manifold $\hat Y$ which obeys
\eqref{dax} and \eqref{dox}
which again will turn out to be responsible for
generating mass terms in the lower-dimensional action.

Let us start by shortly recording the type IIB supergravity in
ten dimensions. The NS-NS sector of the bosonic spectrum consists
of the metric $\hat g_{MN}$, an antisymmetric tensor field $\hat
B_2$ and the dilaton $\hat \phi$. In the RR sector one finds the
0-, 2-, and 4-form potentials $l, \ \hat C_2, \ \hat A_4$. The
four-form potential satisfies a further constraint in that its
field strength $\hat F_5$ is self-dual.
The interactions of the above fields are described by the
ten-dimensional action \cite{JP}
\begin{eqnarray}
  \label{cB1}
  S_{IIB}^{(10)} &=& \int e^{-2\hat\phi} \left(- \frac{1}{2} R *\! {\bf 1} +
    2  d  \hat \phi \wg *  d  \hat\phi - \frac{1}{4}  d  \hat B_2 \wg *  d
    \hat B_2\right) \nn \\[2mm]
  &&-\frac{1}{2}\int \left(  d  l\wg *  d  l+ \hat F_3 \wg * \hat F_3 +
    \frac12 \hat F_5\wg * \hat F_5\right)  \\[2mm]
  && - \frac12 \int \hat A_4 \wg  d \hat B_2\wg  d \hat C_2 \ , \nn
\end{eqnarray}
where the field strengths $\hat F_3$ and $\hat F_5$ are defined as
\begin{eqnarray}
  \label{cB2}
  \hat F_3 & = &  d  \hat C_2 - l  d \hat B_2 \ , \\[2mm]
  \hat F_5 & = &  d  \hat A_4 -  d  \hat B_2 \wg \hat C_2 \nn \ .
\end{eqnarray}
As it is well known the action \eqref{cB1} does not reproduce the correct
dynamics of type IIB supergravity as the
self-duality condition of $\hat F_5$ can not be derived from a variational
principle. Rather this should be imposed by hand in order to obtain the
correct equations of motion and we will come back to this constraint later as
it plays a major role in the following analysis.

In order to compactify the action \eqref{cB1} on a half flat manifold we
proceed as in \cite{GLMW} and
continue to expand the ten dimensional fields in the forms which appear in
\eqref{dax} and \eqref{dox} even though they are not harmonic. We do
not want to go again here through the argument presented in \cite{GLMW}, but
we just mention that the Laplace operator acting on these forms produces
terms of order $($flux$)^2$ and in the supergravity limit, where the fluxes
are supposed to be at a scale much smaller than the compactification one, it
is consistent to keep the massive modes coming from expansion in these
forms and still neglect the massive KK states.
Correspondingly we write
\begin{eqnarray}
  \label{cB4}
  \hat B_2 & = &  B_2 +  b^i \wg \ox_i \ ,\qquad i = 1,\ldots,h^{(1,1)}\ , \\
  \hat C_2 & = & C_2 + c^i \wg \ox_i\ , \nn \\
  \hat A_4 & = & D_2^i \wg \ox_i + \rho_i \wg \tilde \ox^i + V^A \wg \ax_A
  - U_A \wg \bx^A\ ,\qquad A = 0,\ldots,h^{(1,2)} \ , \nn
\end{eqnarray}
and thus one finds the two forms $B_2,C_2,D_2^i$, the vector
fields $V^A,U_A$, and the scalars $b^i,c^i,\rho_i$. Additionally,
from the metric fluctuations on the internal space one obtains
other scalar fields $z^a$ and $v^i$ \eqref{modexp}, which
correspond to the \CY complex structure  and K\"ahler class
deformations respectively. Due to the self-duality condition
which one has to impose on $\hat F_5$, not all the fields listed
above describe physically independent degrees of freedom. Thus as
four dimensional gauge fields one only encounters either $V^A$ or
$U_A$. In the same way, the scalars $\rho_i$ and the two forms
$D_2^i$ are related by Hodge duality and one can eliminate either
of the two in the four dimensional action. In the end one obtains
an $N=2$ supersymmetric spectrum consisting of a gravity multiplet
$(g_{\mu\nu},V^0)$, $h^{(2,1)}$ vector multiplets $(V^a, z^a)$
and $4(h^{(1,1)} +1)$ scalars $\phi, ~h_1,~ h_2,~ l,~ b^i,~ c^i,
~v^i, ~\rho^i$ which form $h^{(1,1)} +1$
hypermultiplets.\footnote{
  We have implicitly assumed that the two-forms $C_2$ and $B_2$ are massless
  in four dimensions and they can be Hodge dualized to scalars  which we have
  denoted $h_1$ and $h_2$ respectively.}

Up to this point everything looks like ordinary \CY
compactification. The difference comes when one inserts the
Ansatz \eqref{cB4} back into the action \eqref{cB1}. Due to
\eqref{dax} and \eqref{dox}, the exterior derivatives of the
fields \eqref{cB4} are going to differ from the standard case
\begin{eqnarray}
  \label{cBhf2}
  d \hat B_2 & = &  d  B_2 +  d  b^i \wg \ox_i +e_i b^i \bx^0 + e_0 \bx^0
  \ , \nn \\[2mm]
  d  \hat C_2 & = &  d  C_2 +  d  c^i\wg \ox_i + e_i c^i \bx^0 \ , \\[2mm]
  d  \hat A_4 & = &  d   D_2^i \wg \ox_i + e_i D_2^i \wg \bx^0 +  d  V^A \wg
  \ax_A -  d  U_A \wg \bx^A + ( d  \rho_i - e_i V^0 )\wg \tox^i \ .\nn
\end{eqnarray}
As in \cite{GLMW} we have also allowed for a normal $H_3$ flux
proportional to $\bx^0$. This naturally combines with the
other fluxes parameters $e_i$ defined in \eqref{dax} to provide
all the $h^{(1,1)}+1$  electric fluxes. With these
expressions one can immediately write the field strengths $F_3$
and $F_5$ from \eqref{cB2}
\begin{eqnarray}
  \label{F35}
  \hat F_3 & = & ( d  C_2 - l  d  B_2) + ( d  c^i - l  d  b^i) \wg \ox_i
  + e_i (c^i - l b^i) \bx^0 - l e_0 \bx^0 \ , \\[2mm]
  \hat F_5 & = & ( d  D_2^i - d  b^i \wg C_2 - c^i  d  B_2) \wg \ox_i +
  (D \rho_i -\cK_{ijk}c^j d  b^k) \wg \tox^i + F^A \wg \ax_A - \tilde G_A \wg
  \bx^A \ , \nn
\end{eqnarray}
where we have defined
\begin{eqnarray}
  \label{cBhf5}
  D \rho_i & = &  d  \rho_i - e_i V^0 \ , \nn \\[2mm]
  F^A & = &  d  V^A \ , \quad G_A =  d  U_A \ , \\[2mm]
  \tilde G_0 & = & G_0 - e_i(D_2^i - b^i C_2) + e_0 C_2 \ ; \quad
  \tilde G_a = G_a \ . \nn
\end{eqnarray}

In order to derive the lower-dimensional action we adopt the
following strategy \cite{GD}. In the first stage we are going to ignore the
self-duality condition which should be imposed on $\hat F_5$ and
treat the fields coming from the expansion of $\hat A_4$ as
independent. Thus, initially we naively insert the expansions
\eqref{cB4} into \eqref{cB1} and perform the integrals over the internal
space using \eqref{normH2}--\eqref{A-N}. To obtain the
correct action we will further add suitable total derivative terms
so that the self-duality conditions appear from a variational principle.
At this point one can
eliminate the redundant fields and in this way obtain the
four-dimensional effective action and no other constraint has to be
imposed. It can be checked that the result
obtained in this way is compatible with the ten dimensional equations
of motion.

Let us apply this procedure step by step. First one inserts the
expansions \eqref{cB4} and \eqref{F35} into the ten-dimensional
action \eqref{cB1}. The various terms of this action take the form
\begin{eqnarray}
  \label{cBhf10}
   - \frac14 \int_Y  d  \hat B_2 \wg *  d  \hat B_2
  & = & - \frac{\cK}{4} \,  d  B_2 \wg *  d  B_2 - \cK g_{ij}  d  b^i \wg *
   d  b^j +\frac14(e_ib^i+e_0)^2 \moo * {\bf 1} \ , \nn \\[5mm]
  - \frac12 \int_Y {\hat {F}_3} \wg * {\hat {F}_3} & = & -
  \frac{\cK}{2}\, ( d  C_2 - l  d  B_2) \wg * ( d  C_2 - l  d  B_2) \nn \\
  & & - 2 \cK g_{ij} ( d  c^i - l  d  b^i) \wg * ( d  c^j - l  d  b^j)
  +\frac12 \Big[e_i(c^i-lb^i)-le_0\Big]^2 \moo * {\bf 1} \ , \nn
\end{eqnarray}
\begin{eqnarray}
  \label{cBhf11}
  -\frac{1}{4}\int \hat F_5\wg * \hat F_5
  & = &  + \frac14 \IM \cM^{-1} \left(\tilde G - \cM F \right)
  \wg * \left(\tilde G - \bar \cM  F\right) \\*
  && -\cK g_{ij} ( d  D_2^i - d  b^i \wg C_2 - c^i  d  B_2) \wg
   *( d  D_2^j - d  b^j \wg C_2 - c^j  d  B_2) \nn \\*
  & & - \frac{1}{16\cK} g^{ij}  (D \rho_i -\cK_{ilm}c^l d  b^m) \wg
   *(D \rho_i -\cK_{jnp}c^n d  b^p)  \ , \nn \\
  & & \nn \\
  - \frac12 \int \hat A_4 \wg d \hat B_2 \wg d \hat C_2 & = &
  - \frac12\cK_{ijk}D_2^i\wg d  b^j\wg d  c^k-\frac12\rho_i\left( d
  B_2 \wg d  c^i +  d  b^i\wg  d  C_2 \right) \ , \nn\\*
  && +\frac12e_iV^0\wg\left( c^i d  B_2-b^i d  C_2\right)-\frac12e_0V^0\wg  d
  C_2 \ . \nn
\end{eqnarray}
In order to write the above formulae we have used \eqref{normH2}--\eqref{A-N}
and we have defined $\moo = \left(\IM \cM^{-1} \right)^{00}$.
In the gravitational sector,
beyond the usual part containing the kinetic terms for the moduli
of $\hat Y$ there will be a further contribution coming entirely
from the internal manifold which is due to the fact that $\hat Y$
is not Ricci flat and which will generate a potential piece in four
dimensions. The Ricci scalar for half-flat manifolds was
computed in \cite{GLMW} and here we will not present the whole
calculation, but just record the effective potential generated in
this way
\begin{equation}
  \label{pot1}
  V_g =  -\frac{\moo}{16\cK}\rme^{2\phi}e_ie_jg^{ij}  \ .
\end{equation}

At this point we have to impose the self-duality condition for
$\hat F_5$ which using \eqref{starox}, \eqref{starax} and \eqref{A-N}
translates into the following constraints on the four dimensional fields
\begin{eqnarray}
  \label{cBhf13}
   d  D_2^i - d  b^i \wg C_2 - c^i  d  B_2 & = & \frac{1}{4 \cK} g^{ij}
  * (D \rho_i -\cK_{ijk}c^j d  b^k) \ , \nn\\[2mm]
  *\tilde G_{A} & = & \RE\cM_{AC} * F^C - \IM \cM_{AC}  F^C \ ,
\end{eqnarray}
with $D \rho_i$ and $\tilde G_A$ defined in \eqref{cBhf5}.
By adding the following total derivative term to the action
\begin{eqnarray}
  \label{cBhf15}
  \cL_{\rm{td}} & = & + \frac12  d  D_2^i \wg  d  \rho_i + \frac12
  F^A \wg G_A \nn\\[2mm]
  & = & + \frac12  d  D_2^i \wg D \rho_i + \frac12 F^A \wg \tilde
  G_A - \frac12 (e_i b^i + e_0) F^0 \wg C_2
\end{eqnarray}
the constraints \eqref{cBhf13} can be found upon variation with
respect to $d  D_2^i$ and $G_A$ respectively. This allows us to
eliminate the fields $ d  D_2^i$ and $G_A$ using their equations
of motion and consequently the effective action obtained in this
way describes the correct dynamics for the remaining fields which
now do not have to satisfy any further constraint.

After the dualization of the 2-forms $C_2$ and $B_2$
to the scalars $h_1$ and $h_2$
one obtains the effective action for type IIB supergravity compactified to
four dimensions on a half-flat manifold
\begin{eqnarray}
  \label{cBhf28}
  S_{IIB}^{(4)} &=& \int - \frac{1}{2} R *\! {\bf 1} - g_{ab} dz^a \wg
  *d\bar{z}^{b} - g_{ij} dt^i \wg *d\bar{t}^j - d\phi \wg *d \phi
  \nonumber\\[2mm]
  && - \frac{e^{2\phi}}{8 \cK} g^{-1\,ij} \left( D\rho_i -
    \cK_{ikl} c^k db^l \right) \wg *\left( D\rho_j -
    \cK_{jmn} c^m db^n \right) \nonumber \\[2mm]
  &&  - 2 \cK e^{2\phi} g_{ij} \left( dc^i - l db^i \right)\wg * \left( dc^j - l db^j \right)
  - \frac{1}{2} \cK e^{2\phi} dl \wg * dl \\[2mm]
  &&  - \frac{1}{2\cK} e^{2\phi}\left(  d  h_1 - b^i D\rho_i +e_0V^0\right)
  \wg *\left(  d  h_1 - b^j D\rho_j +e_0V^0\right) -\rme^{4\phi}D\tilde h\wg *D\tilde h\nonumber \\[2mm]
  && + \frac{1}{2} \RE \cM_{AB} F^A \wg F^B + \frac{1}{2} \IM
  \cM_{AB} F^A \wg * F^B - V_{IIB} * {\bf 1} \ , \nn
\end{eqnarray}
where
\begin{eqnarray}
  \label{cBhf29}
  D \tilde h =  d  h_2 +l  d  h_1 + (c^i-l b^i) D \rho_i + l e_0 V^0 -
  \frac12 \cK_{ijk} c^i c^j  d  b^k \ .
\end{eqnarray}
Performing the field redefinitions \cite{BGHL}
\begin{eqnarray}
  \label{mmap}
  a = 2 h_2 + l h_1 + \rho_i (c^i - l b^i) \ , & & \xi^0 = l \ , \qquad  \xi^i
  = l b^i - c^i \ ,  \\
  \txi_i = \rho_i + \frac{l}{2} \cK_{ijk} b^j b^k - \cK_{ijk} b^j c^k \ ,
  & &
  \txi_0 = - h_1 -\frac{l}{6} \cK_{ijk} b^i b^j b^k + \frac12 \cK_{ijk}
  b^i b^j c^k \ ,\nn
\end{eqnarray}
the metric for the hyperscalars takes the standard quaternionic
form of \cite{FeS} which is now exactly the mirror image of
\eqref{cAns21} with the gauge coupling matrices $\cN$ and $\cM$
exchanged as prescribed by the mirror map. Introducing the
collective notation $q^u = (\phi, a, \xi^I, \txi_I)$ we can write
the final form of the four dimensional action
\begin{eqnarray}
  \label{cBhf33}
  S_{IIA} & = & \int \Big[ -\frac12 R ^* {\bf 1} - g_{ab} dz^a \wg * d
  {\bar z}^b - \tilde h_{uv} D q^u \wg * D q ^v  - V_{IIB}*{\bf 1}\nn \\
  & & \qquad + \frac{1}{2}\, \IM \cM_{AB} F^A\wg * F^B
  + \frac{1}{2} \, \RE \cM_{AB} F^A \wg F^B \Big] \ ,
\end{eqnarray}
where the scalar potential has the form
\begin{eqnarray}
  \label{cBhf34}
  V_{IIB} =  \frac{\moo}{4} \rme^{+ 2 \phi}
  e_I e_J \left(\IM \cN^{-1} \right)^{IJ} - \frac{\moo}{2} \rme^{4 \phi}(e_I
  \xi^I)^2 \ ,
\end{eqnarray}
and the matrix $\cN$ is given in \eqref{eq:N}. The non-trivial covariant
derivatives have the form
\begin{equation}
  \label{cBhf30}
  D\tilde\xi_I =  d \tilde\xi_I-e_IV^0 \ ; \qquad
  Da =  d  a + e_IV^0\xi^I \ ,
\end{equation}
while all the other fields remain neutral.

This ends the derivation of the effective action of type IIB
theory compactified to four dimensions on half-flat manifolds.
One can immediately notice that the gaugings \eqref{cBhf30} are
precisely the same as in the case of type IIA theory
\eqref{cAns07} and \eqref{cAns18} when all the magnetic fluxes
$p^A$ are set to zero.
It is not difficult to see that in this case also the potentials
\eqref{cBhf34} and \eqref{cAns20} coincide.
For this one should
just note that under mirror symmetry $\moo=(\IM\cM_B^{-1})^{00}$
is mapped to $-\frac{1}{\cK_A}$, $\cK_A$ being the volume of the
\CY manifold on which type IIA is compactified.

\section{Conclusions}

In this paper we derived the low energy effective action of type
IIB supergravity compactified on half-flat manifolds and showed
that it is equivalent to the one obtained by compactifying type
IIA theory on \CY three-folds in the presence of electric NS
fluxes. We provided in this way a further check of the conjecture
formulated in \cite{GLMW} that half-flat manifolds represent the
geometry mirror to \CY three-folds with NS three-form fluxes
turned on. However, these half-flat manifolds give rise to only
$h^{(1,1)}$ flux parameters and it seems that one still needs an
additional parameter in order to recover the mirror partners of
all $h^{(1,2)} +1$ electric NS fluxes. Somehow curiously, it was
argued in \cite{GLMW} that this extra parameter arises by turning
on an ordinary NS flux along some particular element of $H^3(\hat
Y)$, $\bx^0$. Here we again found that this prescription works
confirming that this extra flux was not just a coincidence.
Moreover, in analogy to \cite{GVW,gukov} and \cite{GLMW}, we can
write the superpotential
\begin{equation}
  W_B = \int d (B + i J) \wg \Ox \ ,
\end{equation}
which naturally incorporates the additional parameter coming from the flux for
$dB_2$.

We would like to end with an open question which was also posed in
\cite{GLMW}: the magnetic fluxes. The subtlety encountered in \cite{GLMW} was
that in type IIB when
both electric and magnetic NS three-form fluxes were turned on the RR
two-form $C_2$ became massive and the poor understanding of
this issue made it difficult to treat the problem of magnetic fluxes
properly.
However, in the approach we presented in this note, type IIA with electric
and magnetic NS three-form fluxes is well understood and no massive form
is present. Thus, it appears that in this picture it would be easier to
look for the magnetic fluxes and we hope to report on this subject soon
\cite{GLMW2}.

\vspace{1cm}
\appendix
\noindent
{\Large {\bf Appendix}}
\renewcommand{\theequation}{\Alph{section}.\arabic{equation}}

\section{Type IIA with NS flux}\label{IIANS}

In this appendix we briefly recall the results of \cite{LM2} for
the compactification of type IIA supergravity on Calabi-Yau
three-folds $Y$ when background NS fluxes are turned on.

The bosonic spectrum of type IIA supergravity in ten dimensions features the
following fields: the graviton $\hat g_{MN}$, a two-form $\hat B_2$ and the
dilaton $\hat \phi$ in the NS-NS sector and a one form $\hat A_1$ and a
three-form $\hat C_3$ in the RR sector.
The action governing the interactions of these fields can be written as
\cite{JP}
\begin{eqnarray}
  \label{SIIA10}
  S & = & \int \, e^{-2\hat\phi} \left( -\frac12 \hat R *\! {\bf 1} + 2
  d \hat\phi \wg * d \hat\phi - \frac14  \hat H_3\wg * \hat H_3 \right) \nn \\
  & & - \frac12  \int \, \left(\hat  F_2 \wg * \hat F_2 + \hat F_4 \wg * \hat
  F_4 \right) + \frac12 \int \hat H_3 \wedge \hat C_3 \wedge d \hat C_3 \, ,
\end{eqnarray}
where
\begin{equation}
  \label{HFdef}
   \hat H_3 = d\hat B_2\ ,\qquad \hat F_2 = d\hat A_1\ ,\qquad \hat F_4 = d
   \hat C_3 - \hat A_1 \wedge\hat H_3.
\end{equation}
Upon compactification on a \CY three-fold the four-dimensional spectrum can be
read from the expansion of the ten-dimensional fields in the \CY harmonic
forms
\begin{eqnarray}
  \label{fexpA}
  \hat A_1 & = & A^0 \ ,\nn \\
  \hat C_3 & = & C_3 + A^i \wg \ox_i + \xi^A \ax_A + \txi_A \bx^A \ ,\\
  \hat B_2 & = & B_2 + b^i \ox_i \ .\nn
\end{eqnarray}
Correspondingly, in $D=4$ we find a three-form $C_3$, a two-form
$B_2$, the vector fields $(A^0,A^i)$ and the scalars $b^i, \xi^A,
\txi_A$. Together with the K\"ahler class and complex structure
deformations $v^i$ and $z^a$ these fields combine into a gravity
multiplet $(G_{\mu \nu}, A^0)$, $h^{(1,1)}$ vector multiplets
$(A^i, v^i, b^i), \ i = 1,\ldots , h^{(1,1)}$,  $h^{(1,2)}$
hyper-multiplets $(z^a, \xi^a, \txi_a), \ a = 1, \ldots ,
h^{(1,2)}$ and a tensor multiplet $(B_2, \phi, \xi^0, \txi_0)$.

We assume that turning on background fluxes does not change the light spectrum
and thus the only modification in the KK Ansatz is a shift in the field
strength of $\hat B_2$
\begin{equation}
  \label{cAns1}
  \hat H_3 = H_3 + d  b^i\wg\ox_i +p^A\ax_A -q_A\bx^A \ .
\end{equation}
This leads to the following expressions for the different terms
appearing in the ten-dimensional action (\ref{SIIA10})
\begin{eqnarray}
  \label{cAns2}
   - \frac14 \int_Y\hat H_3 \wg * \hat H_3
  & = & - \frac{\cK}{4} \, H_3 \wg * H_3 - \cK
  g_{ij} db^i \wg * db^j - V*{\bf 1} \ , \nn \\ [5mm]
  - \frac12 \int_Y \hat F_2 \wg * \hat F_2 & = &
  - \frac{\cK}{2} \, d A^0 \wg * dA^0 \ , \nn \\ [5mm]
  - \frac12 \int_Y {\hat {F}_4} \wg * {\hat {F}_4} & = & -
  \frac{\cK}{2}\, (d C_3 - A^0\wedge H_3) \wg * (d C_3 - A^0\wg H_3) \\
  & & - 2 \cK g_{ij} (d A^i - A^0 d b^i) \wg * (d A^j - A^0 d b^j) \nn \\
  & & + \frac{1}{2}\left(\IM \cM ^{-1} \right)^{AB}
  \Big[ D\tilde\xi_A +  \cM_{AC} D\xi^C \Big] \wg * \Big[ D\tilde\xi_B +
  \bar \cM_{BD} D\xi^D \Big] \ , \nn \\[5mm]
  \frac12 \int_Y \hat H_3 \wedge \hat C_3 \wedge d \hat C_3
  & = &  - \frac12 H_3 \wg (\xi^A  d \txi_A - \txi_A  d  \xi^A) +\frac12 d b^i
  \wg A^j \wg d A^k \cK_{ijk} \nn \\
  && + d  C_3\wg\left( p^A\tilde\xi_A+q_A\xi^A\right) \nn  \, .
\end{eqnarray}
Even from this stage one can notice that some of
the fields effectively became charged
\begin{equation}
  \label{cAns07}
   D \xi^A =  d  \xi^A - p^A A^0 \quad , \quad D \tilde \xi_A =  d  \tilde
   \xi_A + q_A A^0  \ ,
\end{equation}
and a potential term is induced
\begin{equation}
  \label{cAns06}
  V = -\frac{1}{4}\rme^{-\hat\phi}\left(q-\cM p\right) \IM \cM^{-1} \left(q -
  \bar \cM p \right) \ .
\end{equation}
In order to write the above expressions we have used the following notation
for the integrals on the \CY manifold.\footnote{%
  For a systematic study of the \CY\!\!\! moduli space we refer the reader to
  the literature \cite{BCF,CdO}.}
First the harmonic forms are normalized as
\begin{equation}
  \label{normH2}
  \int_{Y} \ox_i \wg \tilde \ox^j\ =\ \dx_i^j \ ,
\end{equation}
while for $H^3(Y)$ the basis $(\ax_A, \bx^B)$ obeys
\begin{equation}
  \label{icy30}
  \int_Y \ax_A \wg \bx^B = \dx_A^B \ ; \qquad
  \int_Y \ax_A \wg \ax_B = \int_Y \bx^A \wg \bx^B = 0 \ .
\end{equation}
Furthermore we have denoted
\begin{eqnarray}
  \label{K}
  \cK_{ijk} = \int_{Y} \ox^i \wg \ox^j \wg \ox^k  \ , \qquad
  \cK = \frac16 \int_{Y} J \wg J \wg J  \ ,
\end{eqnarray}
where $\cK$ is the volume and $J$ is the K\"ahler form.
Finally, the Hodge duals of the harmonic two-forms are given by
\begin{equation}
  \label{starox}
  * \ox_i = 4 \cK g_{ij} \tox^j \ ,
\end{equation}
where $g_{ij}$ denotes the metric on the moduli space of the K\"ahler
deformations which is given by
\begin{equation}
  4 \cK g_{ij} = \int_Y \ox_i \wg * \ox_j \ .
\end{equation}
For the three-forms we assume the following relations
\begin{equation}
  \label{starax}
  * \ax_A =  {A_A}^B \, \ax_B + B_{AB} \, \bx^B \ , \qquad
  * \bx^A = C^{AB} \, \ax_B - A_B{}^A \, \bx^B\ ,
\end{equation}
where $A, \ B, \ C,$ are given in terms of a matrix $\cM$ which represents the
gauge coupling functions in the case of type IIB
compactification \cite{Suz,CDF}
\begin{eqnarray}
  \label{A-N}
  A & = & \left(\RE \cM \right) \left(\IM \cM \right)^{-1}\ , \nn \\
  B & = & - \left(\IM \cM \right) - \left(\RE \cM \right) \left(\IM
  \cM\right)^{-1} \left(\RE \cM \right)\ ,  \\
  C & = & \left(\IM \cM \right)^{-1} \ . \nn
\end{eqnarray}

Next, the compactification proceeds as usually by dualizing the fields $C_3$
and $B_2$ to a constant and
to a scalar respectively. We do not perform these steps here, but we just
recall the final results. (for more details see \cite{LM2,BGG}). First the
dualization of $C_3$ to a constant $e$ results in
\begin{equation}
  \label{cAns10}
  \cL_e = \cL_{C_3} = -\frac{\rme^{4\phi}}{2\cK}
  \left(p^A \tilde \xi_A + q_A \xi^A + e \right)^2 *{\bf 1}
  + \left(p^A \tilde \xi_A + q_A \xi^A + e \right) A^0 \wg H_3 \ .
\end{equation}
It was shown in \cite{LM2} that the constant $e$ plays a special
role in the case of RR fluxes. however, it is irrelevant for the
analysis in this paper and thus we will set it to zero. Dualizing
now the two-form $B_2$, one obtains an axion, which due to the
Green-Schwarz term in \eqref{cAns10} becomes charged and its
covariant derivative reads
\begin{equation}
  \label{cAns18}
  Da =  d  a - \left(p^A \tilde \xi_A + q_A \xi^A \right) A^0.
\end{equation}

Collecting all terms one can write the final form of the action\footnote{
We have further redefined the gauge fields as $A^i \longrightarrow A^i - b^i
A^0$ and also appropriately rescaled the metric in order to go to the Einstein
frame.}
\begin{eqnarray}
  \label{cAns19}
  S_{IIA} & = & \int \Big[ -\frac12 R ^* {\bf 1} - g_{ij} dt^i \wg * d
  {\bar t}^j - h_{uv} D q^u \wg * D q ^v - V_{IIA}*{\bf 1} \nn \\
  & & \qquad + \frac{1}{2}\, \IM \cN_{IJ} F^I\wg * F^{J}
  + \frac{1}{2} \, \RE \cN_{IJ} F^I \wg F^J \Big] \ ,
\end{eqnarray}
where the potential can be read from \eqref{cAns06} and \eqref{cAns10}
\begin{equation}
  \label{cAns20}
  V_{IIA} = -\frac{1}{4\cK} \rme^{2 \phi} \left(q - \cM p \right) \IM \cM^{-1}
  \left(q - \bar \cM p \right) + \frac{1}{2 \cK} \rme^{4 \phi} \left(p^A
  \tilde \xi_A + q_A \xi^A \right)^2 \ ,
\end{equation}
while the metric for the hyper-scalars $h_{uv}$ has the standard form of
\cite{FeS}
\begin{eqnarray}
  \label{cAns21}
  h_{uv} Dq^u \wg * Dq ^v & = &  d \phi \wg * d\phi + g_{ab} dz^a
  \wg * d \bar z^b \\
  & & + \frac{e^{4\phi}}{4} \, \Big[ Da +
  (\tilde\xi_A D \xi^A-\xi^A D\tilde\xi_A) \Big] \wg *
  \Big[Da + (\tilde\xi_A D \xi^A-\xi^A D \tilde\xi_A) \Big] \nn \\
  & & - \frac{e^{2\phi}}{2}\left(\IM \cM^{-1} \right)^{AB}
  \Big[ D\tilde\xi_A + \cM_{AC} D\xi^C \Big]
  \wg * \Big[D \tilde \xi_B + \bar \cM_{BD} D \xi^D \Big] \ .
   \nn
\end{eqnarray}
Furthermore by $\cN$ we have denoted the gauge couplings matrix which can be
immediately seen from \eqref{cAns2} that it has the usual form \cite{BCF}
\begin{eqnarray}
  \label{eq:N}
  \RE \cN_{00} = - \frac13 \cK_{ijk} b^i b^j b^k \ , \qquad
  \RE \cN_{i0} = \frac12 \cK_{ijk} b^j b^k \ , \qquad
  \RE \cN_{ij} = - \cK_{ijk} b^k \ , \nn \\
  \IM \cN_{00} = -\cK - 4 \cK g_{ij} b^i  b^j \ , \qquad
  \IM \cN_{i0} = 4 \cK g_{ij} b^j \ , \qquad
  \IM \cN_{ij} = -4 \cK g_{ij} \ .
\end{eqnarray}

\vskip 1cm

\subsection*{Acknowledgments}

We would like to thank Jan Louis and Daniel Waldram for encouraging us to
bring this project to an end and for valuable suggestions when reading the
manuscript.

\noindent
This work was supported by the CNRS -- the French Center
for National Scientific Research, DFG -- The German Science Foundation and
DAAD -- the German Academic Exchange Service.


\providecommand{\href}[2]{#2}

\end{document}